# The AAVSO 2011 Demographic and Background Survey

**Aaron Price**
*AAVSO Headquarters, 49 Bay State Road, Cambridge, MA 02138; aaronp@aavso.org*

**Kevin B. Paxson**
*20219 Eden Pines, Spring, TX 77379; kbpaxson@aol.com*

**Abstract**   In 2011, the AAVSO conducted a survey of 615 people who are or were recently active in the organization. The survey included questions about their demographic background and variable star interests. Data are descriptively analyzed and compared with prior surveys. Results show an organization of very highly educated, largely male amateur and professional astronomers distributed across 108 countries. Participants tend to be loyal, with the average time of involvement in the AAVSO reported as 14 years. Most major demographic factors have not changed much over time. However, the average age of new members is increasing. Also, a significant portion of the respondents report being strictly active in a non-observing capacity, reflecting the growing mission of the organization. Motivations of participants are more aligned with scientific contribution than with that reported by other citizen science projects. This may help explain why a third of all respondents are an author or co-author of a paper in an astronomical journal. Finally, there is some evidence that participation in the AAVSO has a greater impact on the respondents' view of their role in astronomy compared to that expected through increasing amateur astronomy experience alone.

## 1. Introduction

   The AAVSO is a large, multinational citizen science organization dating back to 1911. The organization has experienced significant change in the past two decades (Williams and Saladyga 2011), yet our last survey of membership was conducted in 1994. As the organization begins planning for the future, it was time to use data to characterize those who are active and contributing to the AAVSO, in all its forms. This data can be compared with current assumptions and beliefs of the organization and also as a tool for planning new initiatives and direction.

## 2. Prior surveys

   The AAVSO has conducted a number of membership surveys in the past 35 years. These surveys included a mix of demographic (such as "age"), programmatic (such as "observing trends") and operational (such as "evaluation of staff") items.

In 1976, AAVSO Director Janet Mattei, with staff members Linda M. Blizzard, and Joseph M. Manella, mailed a survey to members. The survey included sections on headquarters operations, communications, observations, meetings, and publications. Approximately 200–300 responses were received and are stored in the AAVSO archives (AAVSO 1976). However, no known tabulation or report of the responses is known to exist.

In January of 1980, the AAVSO mailed a second survey to members. It was a smaller survey, with ten questions focused on demographics and observing activities. A summary of results were reported in Waagen (1980). 267 surveys were returned as of the writing of that report. None of the questions on the 1980 survey were also on the 1976 survey, so the two surveys could be seen as complimentary.

In 1994, Wayne M. Lowder designed a survey of members and observers in response to a request by a recently convened Futures Studies Group. The survey had 79 items divided into sections focused on demographics, publications, other resources of information, astronomical activities, observing, headquarters operations, meetings, and use of personal computers. 420 surveys were returned and tabulated by AAVSO staff (Tonja Foulds, Shawna Helleur, Dennis Milon, and Barbara Silva). Results were presented to the Futures Studies Group in the form of an executive summary written by Lowder in September 1994, which exists in the AAVSO archives (Lowder 1994). The Futures Study Group presented results to the AAVSO Council at the AAVSO Annual Meeting in October, 1994 (Hazen 1994).

The AAVSO has also run a few surveys over the past few years focused on more specific topics. In 2010, the AAVSO conducted a survey with eight questions about AAVSO meeting experiences. This survey was distributed exclusively online through the AAVSO website and received 88 responses. Since 2009, the AAVSO's Citizen Sky project has asked participants a few optional demographic questions when they first registered for the Citizen Sky website. 1,385 of these responses were analyzed in a paper by Price and Lee (in press) and some of that data is included here.

**3. Survey design and methodology**

The goal of the 2011 survey was to better characterize the AAVSO membership so that staff and the Council can make better decisions regarding membership activities and future directions of the organization. This includes testing current assumptions about members and also looking for unexpected results in the data. As such, the survey items were designed to report on the respondents' educational and professional backgrounds and their experience in the AAVSO and astronomy in general.

The survey was designed prior to the known existence of the previous surveys. Yet the items on the surveys are often quite similar, which we feel is a testament to the validity of the chosen items. Twenty of those who were privately invited to take the survey tested the first draft. Only technical changes came out of that pilot test. The survey has a maximum of 27 items (see Appendix A). However, some items are conditioned only to appear based on certain responses to earlier items and all items were optional. So the response rate varies item-to-item.

The survey was placed on the Survey Monkey website so that results could be automatically tabulated. We posted the survey's URL to the AAVSO Discussion Group (~490 subscribers) and on the AAVSO web site (417 reads). We sent an e-mail to those who were current AAVSO members or who had made an observation within the last five years (~2,400 e-mails). We identified eight people who met that criteria but who did not have e-mail addresses on record. For them, we printed a copy of the survey and sent it using postal mail. A total of 691 valid responses were received to the online survey and four of the printed surveys were returned.

We tabulated the results into an Excel spreadsheet. For the open-ended items, we coded them into a set of categories that included 99% or more of the responses. When a particular item response could fit into more than one category, the first category mentioned in the response was used. This was based on the assumption that the first item mentioned by the respondent was the most important to them. The ranking items were scored on an ordinal scale (the highest ranking item is assigned a "1" and the rest are ranked accordingly). We treated items that were not answered as missing data. We made our final analysis with the PASW Statistics 18 software (formerly SPSS Statistics, now IBM SPSS Statistics).

3.1. Definitions

For the purposes of this survey, we refer to *respondents* as anyone who answered at least one item of the survey. Also, we combined Charge-Coupled Device (CCD), Photoelectric Photometry (PEP), and Digital Single-Lens Reflex (DSLR) technologies beneath the umbrella term of *digital technologies*. *Visual observations* include any observation made with the eye, which includes naked eye, binoculars, and telescopic observations made with an eyepiece. *Membership status* was assigned based on AAVSO headquarters records as of January, 2012.

The profession categories included items were taken from the U.S. Department of Labor categories used in the 2011 U.S. Census (USDOL 2011). The categories of objects were taken from the highest level categories used by the Variable Star Index (VSX) (AAVSO 2011), which are based on categories originally developed by the *General Catalog of Variable Stars* (GCVS; Kholopov *et al.* 1984).

## 4. Results

4.1. Age

The mean age of survey respondents is 53 (N=671; Figure 1). There is no significant difference between the mean ages of men and women, nor of the mean ages between digital and visual observers. The 1994 survey reports frequencies instead of means. The mean frequency category was "41–50" (N=420). Citizen Sky members report a mean age of 41 (N=1,385). *Sky & Telescope* magazine reports a mean subscriber age of 51 (New Track Media 2010).

The AAVSO maintains an archive of membership applications dating back to 1911. Almost all applications include either the applicant's age or birthdate. We randomly selected 615 applications and plotted their age as of the moment they joined the organization (Figure 2). There are some gaps in the data from incomplete records (namely 1911–1918, 1922–1928 and 2004–2008). Over the entire 100-year period, the average age of an AAVSO membership applicant was 37 years old. During 1911–1921, the average new member age was 40 years. New member age dropped to 28 years during the years 1967–1977 and was 51 years for the years 2001–2011.

The average age of observers does not vary much between the various observing techniques included in the survey (Figure 3). The age of PEP observers is the only category that stands out. We found slightly more variation between the average ages of those engaged in non-observing activities (Figure 4). In particular, the more "high-tech" activities of programming and data mining tend to have slightly younger participants.

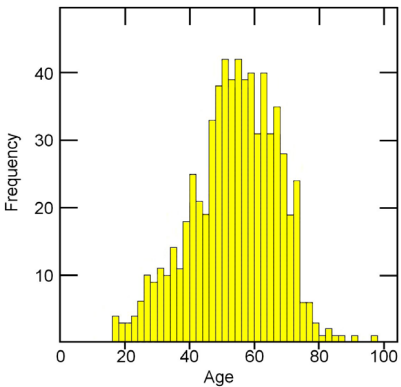
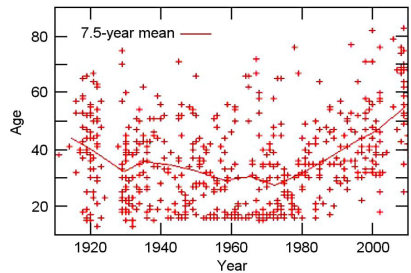

Figure 1. Histogram of age of survey respondents. The mean age is 53.31; Standard Deviation = 13.384; N = 671.

Figure 2. Age of new member applicants.

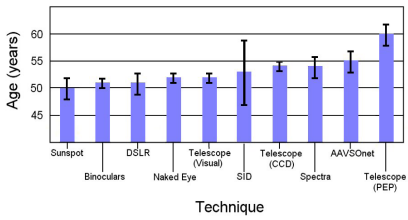

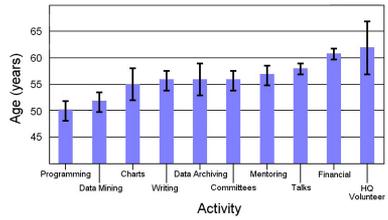

Figure 3. Average age of survey participants as a function of observing technique.

Figure 4. Average age of survey participants as a function of non-observing activity.

4.2 Years active

"Years active" is a variable computed from subtracting the first year a respondent reported to be active in the AAVSO from 2011. It represents an approximation of how long survey respondents have been active in the organization. The mean is 14.3 years (SD=15, N=598). There seems to be a drop off at years 2 and 5, after which drop out rates flatten (Figure 5). Members of the AAVSO tend to be involved in the organization for six years longer than non members, a difference which is statistically significant, $F (1,503)=22.0$, $p<.001$.

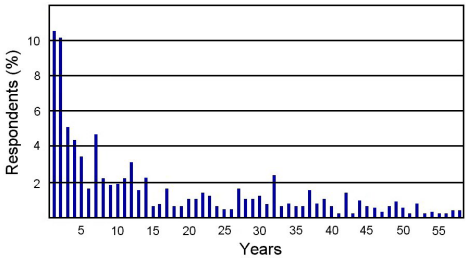

Figure 5. Number of years active of all 2011 survey respondents.

The 1994 survey included frequencies of the number of years respondents have observed variable stars (note a slight difference between their question about "observing" and our question about "being active") separated into 5-year bins (Figure 6), so we were unable to compute a mean. The only major difference between the 1994 and 2011 distributions is a drop off after 30 years that appears in the 1994 survey, but not in the 2011 survey.

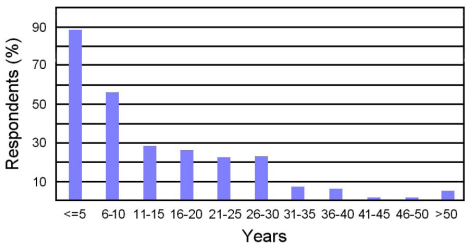

Figure 6. Number of years active of observers only, from 1994 survey.

In order to get more detail from the period around the 1980 survey, we randomly selected 26 observers from the AAVSO International Database (AID) who had submitted observations in 1977 (chosen to match the same interval between 1994 and 2011) and pulled their original membership application to set a date for their joining of the organization. We computed the difference between that date and 1977 as their AAVSO Age (Figure 7). The mean age was 10.8 years. The mean AAVSO Age of members in the 2011 survey was 16.8 years. The distributions are similar, but not the same. The drop-offs seen in the 2011 survey at years 2 and 5 occur in the 1977 data, but a year or two later. Overall, the trends are very similar.

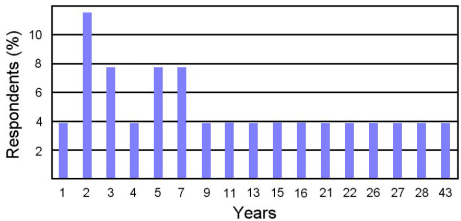

Figure 7. Years of activity of members and observers active in 1977 (N = 26).

4.3 Membership status

Membership status was determined by looking up an observer code or e-mail address (when provided) in the AAVSO membership database on January 18, 2012. At that time, 54% of respondents were official members of the AAVSO. In the 1994 survey, 85% of the respondents were members of the organization (N=417), according to self reported data. In the 1994 survey, we received observations from 660 observers. In 2011, we received observations from 1,050 observers. The observer/membership ratio difference may reflect more the growth of observers rather than the loss of members. The AAVSO does not keep a record of membership totals per year.

We also looked for a relationship between membership status and whether the respondent reports to be an active observer or not. We found no significant relationship. However, for those who were active, there was a significant relationship between the techniques they used and their membership status, $F(390,8)=2.21$, $p=.03$ (Figure 8). The most interesting result is that 60% of telescopic CCD observers (N=143) are members and 43% of telescopic visual observers (N=138) are members. This difference is statistically significant, $F(315,1)=7.36$, $p<.01$.

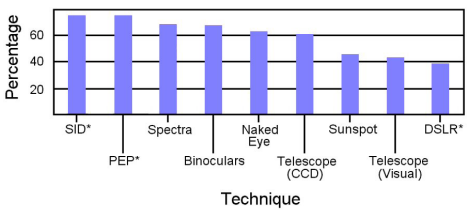

Figure 8. Membership rates for particpants as a function of various observing techniques (* = categories with fewer than 15 member respondents).

4.4. Gender

92% of respondents identified as male (N=634) and 8% identified as women (N=44). The mean age for women is 49, while the mean age for men is 53, however the difference is not statistically significant (p=.06). The low sample of females makes it difficult to look for relationships between gender and other variables in the survey. In the 1994 survey, the distribution was 94% male and 6% female, very close to the current ratio. *Sky & Telescope* reports a gender ratio of 95% male and 5% female in their 2010 advertising rate card (New Track Media 2010). The Citizen Sky gender distribution is 78% male, 19% female and 3% unreported (N=1,385).

4.5 Country

108 different countries were represented in our survey results. About 49% of respondents were from the United States. The rest were widely distributed among the other 107 countries (Figure 9). 28 countries were represented by members and 46 countries were represented by active observers.

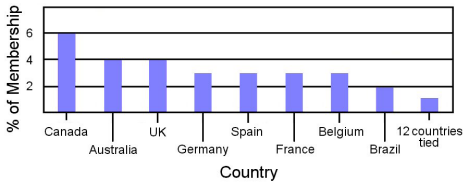

Figure 9. Top countries represented by membership, excluding the United States.

4.6 Formal education

Almost a quarter of the respondents claimed to have a terminal degree (Ph.D., M.D., J.D., and so on) in their field (Figure 10). About 76% report a Bachelor's degree or higher, which is close to *Sky & Telescope's* rate of 77% (New Track Media 2010).

There is a significant relationship between profession and observation type, F(9,438)=3.44, p<.01. Most of that significance is due to the increased education levels of the spectroscopic observers (N=32) and decreased education levels reported by the sunspot observers (N=22) (Figure 11).

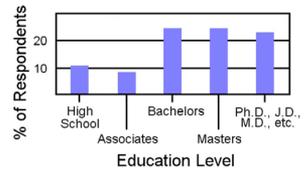

Figure 10. Reported formal education of respondents (N = 656).

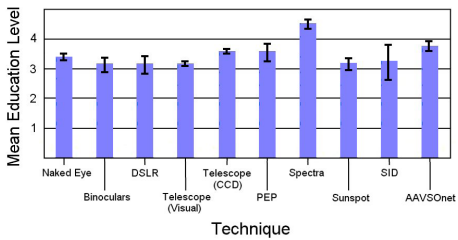

Figure 11. Average formal education of the users of various observation techniques. Education level reflects the categories in Figure 10 in ascending order.

There are also some correlations between education and the type of objects the respondents favored to observe (Table 1). Specifically, those with less of a formal education tend to be more interested in novae (N=421, q=.139, p < .001) and the Sun (N=404, q=.121, p < .05). Also, those with more of a formal education tend to be more interested in rotating variables (N=354, q=−.107, p < .05). All of these relationships are considered small by traditional social science standards (Cohen's Guidelines).

Table 1. Intercorrelations between formal education and interest in objects.

| CV | EB | Extra-galactic | Novae | Non-Stellar | Pulsating | Rotating | Sun | YSO |
|---|---|---|---|---|---|---|---|---|
| | | | | Formal Education | | | | |
| −.044 | −.055 | .074 | .139* | .022 | −.055 | −.107** | .121** | −.049 |

*$p < .01$, **$p < .05$.

4.7. Profession

The reported distribution of professions (N=615) can be broken down into two categories of high and low (Figure 12). The most common professions were in science, computer science, engineering, and education. Those four categories account for about 57% of the respondents. The rest of the other categories were roughly even, with management and health care leading the group. Building and Grounds Cleaning had the fewest responses.

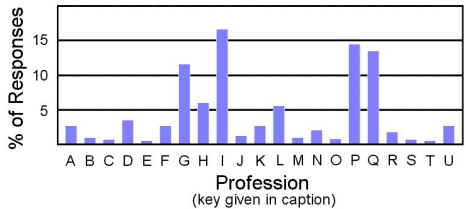

Figure 12. Distribution of professions according to U.S. Department of Labor categories: a. Transportation and material; b. Sales and related; c. Protective service; d. Production; e. Personal care and service; f. Office and administrative support; g. Mathematical and computer; h. Management; i. Life, physical, and social science; j. Legal; k. Installation, maintenance and repair; l. Healthcare practitioners and technical; m. Food preparation and serving; n. Financial; o. Farming, fishing, and forestry; p. Engineering, architecture and surveyors; q. Education, training, and library; r. Construction and extraction; s. Community and social services; t. Building and grounds cleaning; u. Arts, design, entertainment.

The 1994 survey had a similar question, but with fewer profession categories to choose from. Table 2 is a comparison of results from the two surveys in categories that are similar. In general, there is not much difference that cannot be explained by differences in the definition/labeling of categories between the surveys.

Table 2. Comparison of 1994 and 2011 reported professions (selected).

| 1994 Category | 2011 Category | 1994 Response (%) | 2011 Response (%) |
|---|---|---|---|
| Scientific/Technical | Life, Physical, and Social Science | 41 | 17 (31*) |
| Professional Astronomer | (none) | 12 | 13** |
| Educator | Education, Training, and Library | 18 | 14 |
| Computer Specialist | Mathematical and Computer Scientists | 9 | 11 |
| Business Management | Management | 5 | 6 |
| Sales/Marketing | Sales, and Related | 1 | 1 |

*When combining this category with "Engineering, Architecture, and Surveyors."
**Taken from astronomy experience responses (see text).

4.8. Astronomy experience

Almost half (49%) of the respondents classified themselves as having an advanced level of experience in astronomy (Figure 13). There is a significant, but low positive correlation between astronomy experience and the variable we call "Years Active" ($r = .208$, $p < .001$). There is also a significant correlation between age and astronomy experience, $r = .087$, $p < .05$, but it is weaker than the relationship with Years Active. In fact, when controlling for Years Active through an ANCOVA, age is no longer a significant predictor of experience, $F(67, 587) = 1.11$, $p = .125$.

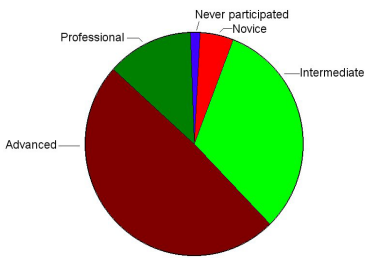

Figure 13. Astronomy experience levels of of respondents (N = 658).

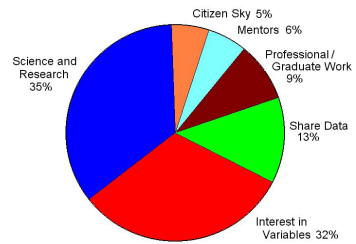

Figure 14. Sources of motivation to participate in the AAVSO (N = 351).

4.9. Motivation

Supporting science and research ("citizen science") is the most popular reason respondents gave for being active in the AAVSO. A close second is an interest in variable stars (Figure 14). Around 9% of respondents are active in the AAVSO due to professional or graduate student research.

4.10. Barriers to activity

To investigate barriers to activity we asked an open-ended question worded as: "If you are not currently active in the AAVSO, what is the main reason?" Almost all respondents who previously reported to be inactive (N = 192) provided an answer to this item. Note the vague definition of "active" in the question. We purposely allowed the respondent to define activity in their own way so as to include those who would otherwise be active in non-observing contexts. Time (43%) was by far the main reported reason respondents were inactive (Figure 15). Other astronomy interests was second (14%), followed by a lack of equipment/poor location (12%). Much of poor location comments were reported as problems with light pollution.

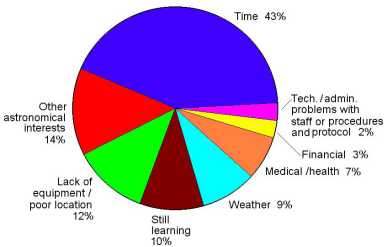

Figure 15. Major reasons to be inactive / barriers to activity (N = 192).

4.11. Referral sources

People learn about the AAVSO from a variety of sources (Figure 16). Word of mouth (25%) is most common, followed closely by other astronomy club or conferences (19%), *Sky & Telescope* magazine (19%), and the Internet (18%). The 1994 survey also included a question about referral sources. It was stated as: "How did you first hear about the AAVSO?" and had 427 responses (Figure 17).

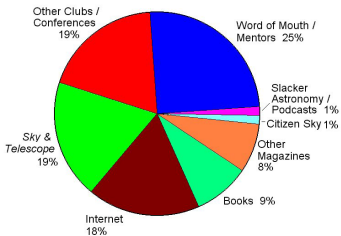 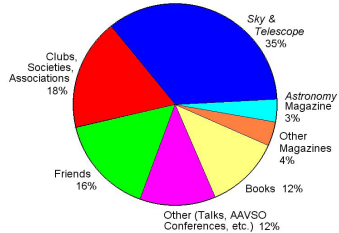

Figure 16. Referral sources in the 2011 survey.  Figure 17. Referral sources in the 1994 survey.

The biggest difference between the 1994 and 2011 surveys is the 35% to 19% drop in referral share attributed to *Sky & Telescope* magazine. *Sky & Telescope* has long been an important source of branding for the organization, however it has dropped in significance. This is likely due to the ability of the AAVSO to reach amateur astronomers directly through the Internet. If you add up the referral share of the "Internet" and "Sky & Telescope" in this survey it reaches 37%, which is very close to the 35% share in the 1995 survey (and in line with general societal trends). Club and books continue to have similar shares between the 1994 and 2011 surveys, at 18%–19% and 12%–9% respectively. Interestingly, non-*Sky & Telescope* magazines also have a similar share between surveys, at 7%–8%.

In the 2011 survey, we coded talks into the "Word of mouth" category. If you combine the "Other (talks…)" and "Friends" categories in the 1994 survey, then they too have a similar share with the 2011 "Word of mouth" category at 28%–25%.

The 1980 survey included a referral question as well. It was stated as: "Source of information about the AAVSO?". Waagen (1980) divided the responses into 5 categories (Figure 18). In general, they are consistent with the other surveys. The major difference being the large share books held in 1980 (21%) as opposed to 1994 (12%) and 2011 (9%). Club referrals also dropped between 1980 (23%) and 1994 (18%) while remaining consistent from 1994 to 2011 (19%).

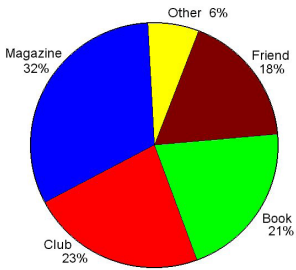

Figure 18. Referral sources in the 1980 survey.

4.12. Observation activity

We asked respondents to declare whether they consider themselves active observers or not. We did not define the term "active." Almost 2/3rds (63%) of the respondents report to be active (Figure 19). Similarly, the 1994 survey asked observers to classify themselves into various categories of activity. 78% chose a category denoting activity and 22% chose a category denoting inactivity. Thus, the percentage of inactive observers has increased since the last survey. In the 1980 survey, 93% reported to be "an observer."

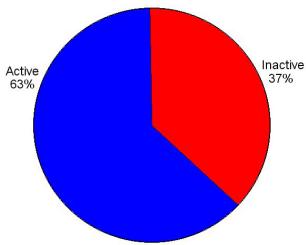

Figure 19. Active and inactive observer rates (N = 691).

We were interested to see whether the increase in inactive observers was related to any of the non-observing activities in the AAVSO. An ANOVA found a significant difference between the active and inactive groups based on whether they are also active in a non-observing activity, $F(10,644) = 1.9$, $p < .05$. Figure 20 is a plot of the mean observation activity value (lower number means more active) grouped by non-observing activities listed in this survey. It is of no surprise that the many active observers are those involved in the chart process, since charts have a direct impact on observing. It is also interesting that the least active are those involve in financial aspects of the organization. Programmers tend to be active observers as well. Beyond that, the rest of the categories are roughly even.

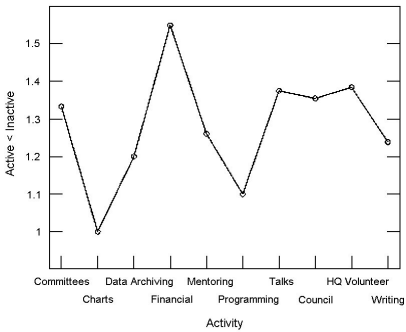

Figure 20. Activity rates of respondents who participate in non-observing activities.

4.13. Faintest observation

We asked observers to report the magnitude of the faintest observation they "typically" observe (Figure 21). We then asked whether that observation was made visually or with "CCD/DSLR/PEP/Other Digital system." This was more useful than asking if a person was a "visual or CCD" observer since many use both techniques. Instead, this question tells us which technique they used to get their faintest observation, which will almost always be "CCD" for CCD/combined observers and "visual" for visual-only observers (for a more detailed discussion of observation type see the observation technology section). The goal of these questions was to establish magnitude ranges for visual or CCD campaigns. Observers were divided roughly in half between the two (Figure 22). The mean faintest visual observation was 12.02 and the mean faintest CCD observation was 13.75. The visual distribution drops off sharply around magnitude 15–16 while the CCD distribution fades more gradually until around magnitude 20 (Figure 23).

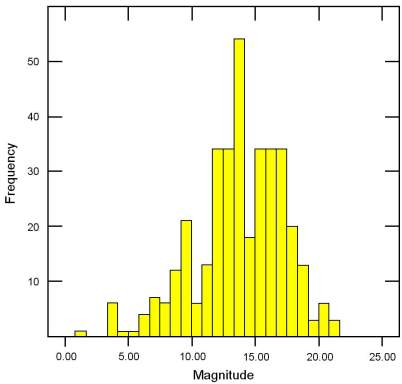

Figure 21. Faintest observations (magnitude) "typically" recorded by respondents. Mean = 13.75; Standard Deviation = 3.45; N = 365.

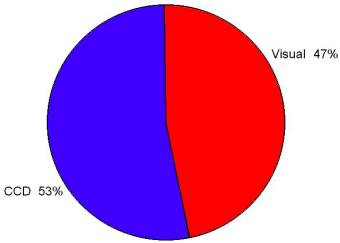

Figure 22. How the faintest observation in Figure 21 is measured.

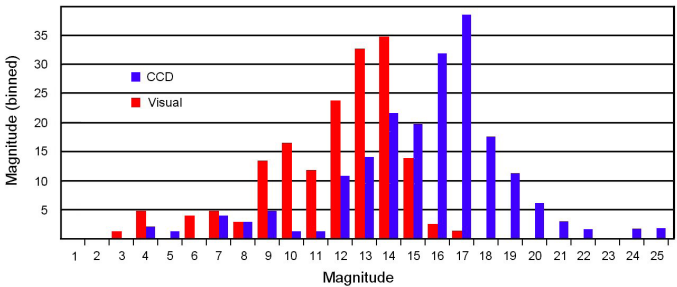

Figure 23. Distribution of the faintest "typical" observation made through CCD or visual measurements.

4.14. CCD origins

We were interested in how often visual observing is used as a stepping stone to digital observing. We asked digital observers: "If you are a CCD/DSLR/PEP observer, did you begin as a visual observer who migrated to CCD/DSLR/PEP or did you begin as a CCD/DSLR/PEP observer?" About half of the respondents report beginning as a visual observer (Figure 24).

This result is somewhat biased because it includes the generation of AAVSO observers who began when visual was the only option. When we looked at the answers of only those who became active within the last 10 years (using the "Years Active" variable, N = 119) we found 67% began with CCD and only 33% began as visual observers. If we tighten the window further and only look at the last 5 years (N = 79), we don't find much additional difference as about 28% of CCD observers began as visual observers (Figure 25). Currently, visual observing is an effective entry point for roughly a quarter of new AAVSO CCD observers.

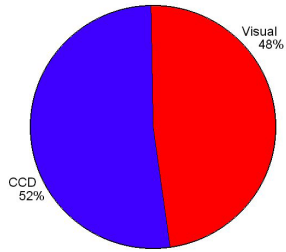

Figure 24. Background (origins) of active CCD observers.

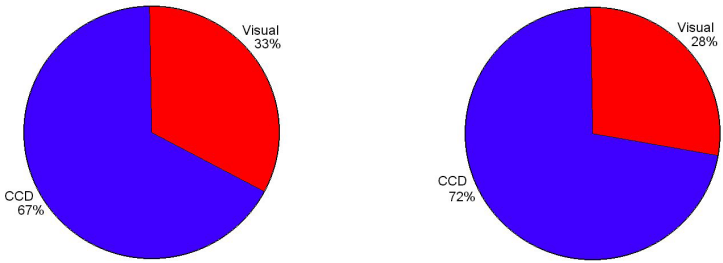

Figure 25. Backround (origins) of active CCD observers for the last 10 years (left), and last 5 years (right).

### 4.15. Observation technology

We provided respondents with a list of observation technologies and asked to select which types they "actively use" (N=460, Figure 26). Interestingly, 40% of those who reported to be telescopic CCD observers are also active as visual telescopic (N=54) or naked eye (N=34) observers. Also, the number of those who reported to be observers utilizing spectra (N=32) was surprising, since the organization has no formal group or section to standardize this activity. Lastly, more than half of all observers use multiple observing techniques (Figure 27).

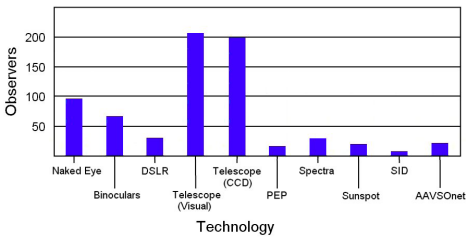

Figure 26. Number of active observers using various observation technologies.

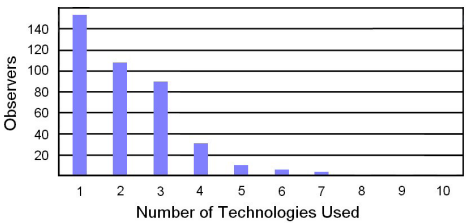

Figure 27. Number of different observing technologies used by active observers.

### 4.16. Non-observing activities

By far, the most popular non-observing activities involved public outreach (Figure 28). Specifically, giving talks and writing about the AAVSO. Prior surveys did not include items to address participation in these areas. We found no significant relationships between non-observing activity and observation technology or with membership status.

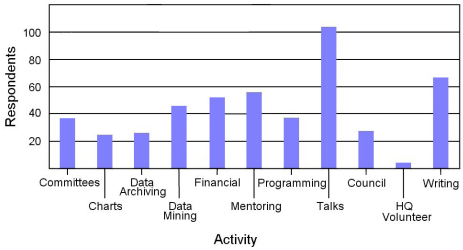

Figure 28. Number of respondents participating in various non-observation activities.

### 4.17. Object interest

When asked to rank objects by interest, we find a clustering of objects into two categories: highly ranked (Pulsating, CV, EB, Novae, Extragalactic) and lesser ranked (YSO, Sun, Rotating, and Non Stellar) (Figure 29). Respondents were only allowed to provide one object for each ranking. However, they were not required to rank every object. Some respondents only ranked the objects they were most interested in (example: they only ranked the top 5). In a separate tabulation, we assigned a ranking of 9 (the lowest rank possible) to those objects that had missing data, with the justification that the respondent had no interest in that object. The only major difference is that the drop off between the highly ranked group and the lesser ranked group increases. 33 types of objects were included in the "Other" category (Table 3). Many of these types of objects could be included in existing survey categories.

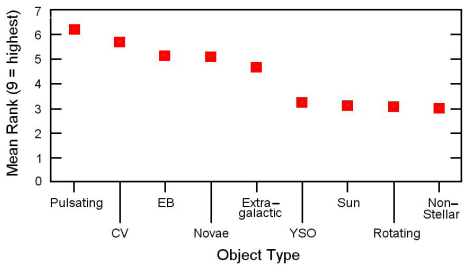

Figure 29. Popularity of various variable star object types.

### 4.18. Meeting attendance

The mean number of meetings attended by a respondent was 1.5 (N = 584; Figure 30). However, that number is significantly skewed because of four respondents of reported between 20–60 meetings. The overall survey median value is 0, meaning that the vast majority of respondents had not attended an AAVSO meeting before. By excluding those who have not attended any meetings, and the two persons who reported 40 and 60 meetings, then the average meeting attendance is 4.2. This number reflects the number of meetings someone would typically attend if they have attended at least one.

Table 3. Objects included in the "Other" field.

| "Other" Objects | Number | "Other" Objects | Number |
| --- | --- | --- | --- |
| Exoplanets | 17 | HAD's | 1 |
| Asteroids | 9 | Supernovae | 1 |
| RCB's | 6 | Cool Supergiants | 1 |
| Be, Shell or GCAS stars | 4 | Central-planetary nebulae | 1 |
| Symbiotic variables | 3 | Thermal pulse candidates | 1 |
| Carbon | 3 | AGB's | 1 |
| Spotted or RS CVn | 2 | Emission line stars | 1 |
| GRB's | 2 | Mira's | 1 |
| HMXB's | 2 | NSV's | 1 |
| Globular variables | 2 | ISM | 1 |
| Non-specific (general) | 2 | RR Lyr | 1 |
| Bright variables | 2 | Del Sct | 1 |
| Novae recurrent | 2 | SX Phe | 1 |
| SR's | 2 | Neutron stars | 1 |
| Irregulars | 1 | White dwarfs | 1 |
| Double or multiple | 1 | Interesting stellar spectra | 1 |
| Infra-red stars | 1 | | |

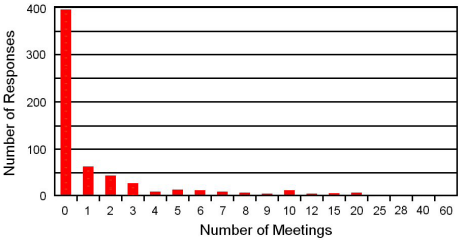

Figure 30. Number of AAVSO meetings attended by respondents.

We looked for relationships between meeting attendance and the type of observations people make and the types of non-observing activities they participate in (Figure 31). For types of observations, we found no significant relationships. For non-observing activities, we did find a significant relationship, $F(573, 10) = 15.4$, $p < .001$. However, most of that statistical relationship is due to those who selected "HQ Volunteer." We believe this to be a perplexing variable because those who volunteer at Headquarters tend to live near Headquarters, thus attend the annual meeting quite often. In Figure 31, we zoomed in on the other categories. The main difference between the activities that involve more investment of initiative (writing, programming, mentoring, and so on) than in activities that are more procedural based. Finally, members are more likely to attend meetings (mean = 2.5 meetings) than nonmembers (mean = 0.3 meetings). This difference is statistically significant, $F(1, 483) = 31.1$, $p < .001$.

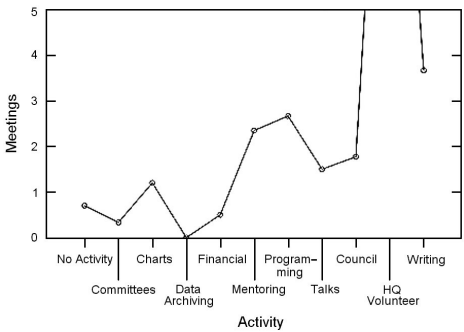

Figure 31. Number of meetings attended by participants of various non-observing activities.

4.19. Paper authorship

A little over a third of the respondents reported to be either an author or coauthor of a paper in an astronomical journal (36%). None of the prior surveys included paper authorship.

## 5. Discussion

Overall, the 2011 survey depicts those participating in AAVSO activities as similar to those described in the 1980 and 1994 surveys. They are largely male, older, highly educated and tend to work in scientific or technical fields. Most are active observers and have been affiliated with the AAVSO for decades.

The survey also suggests areas of significant change over time. First, the average age of new members has been increasing. This began in the 1980s and has continued. Originally, the increase in age was mostly attributed to the loss of younger members. More recently it can also be attributed to an increase in older members. The overall size of the organization's membership has not diminished over this same time period. This "greying of astronomy" is an issue that affects amateur astronomy as a whole. The average age of a *Sky & Telescope* subscriber was 39 in 1979, 48 in 1998, and 51 in 2010 (Beatty 2000; New Track Media 2010). An analysis of the average age of professional American astronomers found an increase of around half a year per year in the mid 1980s (Thronson and Lindstedt 1986). However, the AAVSO Citizen Sky project has been more successful in recruiting younger observers. It reports a mean age of 41 (N = 1,385). For comparison, the median age in the United States as of the 2010 census was 37. The cause behind this trend is complicated and multifaceted. One issue could be the impact of the cold war space race on interest in science in the 1950s and 1960s. Another could be the increasing entertainment options for a decreasing amount of personal time. Amateur astronomy has also become a largely high-tech endeavor requiring significant initial financial investment.

A more detailed investigation of the age question is planned for a future study. Another major difference between surveys is where and how respondents heard about the AAVSO. The Internet is replacing many of the referrals which previously may have come through *Sky & Telescope* magazine. It is interesting that referrals from books and non-*Sky & Telescope* magazines have not changed much between the 1994 and 2011 surveys. This suggests the issue may be specific to *Sky & Telescope*, perhaps due to its previously dominate position as a major referring source and/or because *Sky & Telescope* is more closely associated with advanced amateur astronomers than casual readers. It is possible that the Internet is impacting newsstand sales (casual readers) less than circulation sales (more advanced readers).

In terms of observing methodology, active observers are pretty evenly divided between digital and visual observing. Since the 1990s, there has been discussion about competition between the two types of observing. However, we found that 40% of telescopic CCD observers are also visual observers. Also, about half of all CCD observers and a quarter of new CCD observers began as visual observers. This suggests the line distinguishing these two groups is much fuzzier than has been advertised.

There are some surprising results as well. First, AAVSO activity is related to a greater increase in self-efficacy in astronomy than one would find through increased astronomy experience alone (as measured through age). That is, it is possible that the demands of variable star research have a greater impact on how one views their knowledge of astronomy than the activity of the typical amateur astronomer. This could hint at greater learning taking place in active citizen science projects when compared to typical amateur astronomy projects. A future study is planned to investigate this result. A second surprise was the number of respondents who report to be active in the AAVSO, but also report to be inactive observers. This reflects the increased scope of the AAVSO over the past two decades. When the past two surveys were conducted, non-observing activities were not even considered unless they were in support of observing. Now, many participants of the AAVSO are dedicated to important projects such as programming, analysis of data, public outreach, and so on. Observing is still the heart of the AAVSO, with 60% of those who report participation in at least one non-observing activity also report to be active observers. Another surprise was the high level of education reported. Over half of participants report a graduate degree, with 24% reporting a terminal degree in their field (Ph.D., M.D., J.D., and so on). Also, 13% identify as professional astronomers (N=86). One of the most interesting surprises is the number of countries represented by AAVSO membership (108). This may reflect the significant work that volunteers have put into translating our training materials into other languages, but it is likely a bigger reflection of the universal appeal of variable stars. There are simply so many types of stars and so many open questions, that almost anyone can find a project or object of interest. Finally, the number of respondents who have authored or coauthored a paper in a scientific journal was quite high (36%). This is one of the major distinguishing characteristics between the AAVSO and other citizen science organizations, who tend to focus on using participants to contribute data for professionals to analyze and publish.

Raddick, *et al.* (2010) identified 12 categories of motivation from Galaxy Zoo participants through analysis of interviews and online forum posts. As with our survey, their categories were reduced from open ended discussion (in their case, interviews were also included). In areas where our categories overlap with theirs, we find similarities and some significant differences in motivation rates. They report only 1% of their responses are motivated by "science" while "science and research" was the motivation of 35% of our respondents, which was our highest category of motivation. The Galaxy Zoo primary motivation category was "astronomy" at 39%. If that refers to interest in astronomy, then it is analogous to our "interest in variable stars" category that was cited by 32% of our respondents. Our two groups report the same level of motivation in terms of contributing data to a greater cause. 13% of the Galaxy Zoo participants cite a desire to "contribute" as a motivation of their participation while an identical 13% of our respondents cite a desire to "share data."

This study has a number of limitations. First, it is a study of active or recently active participants of the AAVSO. Some data, such as the ranked interest in types of objects, will be skewed towards current operations (people interested in exoplanets, for example, may have dropped out of the organization). So this data should not be used as a guide for the future, but only as a snapshot of the present. Second, the coding of the open-ended items was limited to one code per item. So it may oversimplify the results of those items. There is a striking similarity between our results and past surveys on these items, which suggests strong validity. Finally, this is self reported data. Thus it includes biases caused by human nature and different definitions of terminology. For example, some respondents report to be professional astronomers yet also report to not have a Ph.D. We are not stating they are not professionals, just that respondents will have difference definitions of the term "professional." To some, it requires a "Ph.D." while for others it denotes publishing in journals while still others apply the term to anyone contributing scientifically to astronomical research at any level.

## 6. Conclusion

This is a summary report of the AAVSO 2011 Demographic Survey, which included current and recent AAVSO participants. Compared with past surveys of this type, it shows an organization that is largely similar in demographics. Respondents were active in a wide variety of observing and non-observing activities and are interested in a wide variety of objects. The AAVSO reflects a "big tent" mentality, with room for everyone interested in variable stars. There are signs in the data of some challenges, such as a population that is growing older and the presence of a very significant gender gap. However, these are not limited to the AAVSO alone. As a descriptive analysis, we make no predictions for the future. However, the results can be used, along with other surveys and analysis, to identify future paths and opportunities for the organization.

# Appendix A

*This is the printed version of the AAVSO 2011 Demographic survey. All items are worded exactly as they appeared online, except for the state and country items, which included drop down lists. In a few areas, screen shots of the online form were used in the printed survey as well.*

**AAVSO 2011 Demographic survey**

1. What is your name? _______________________________________________
2. If you have an observer code, what is it? _____________________________
3. Do you want a copy of the summarized results of this survey?
    If so, enter your email address here: _______________________________
4. What is your gender? ______________________________________________
5. What is your age? _________________________________________________
6. Zip code or postal code: ___________________________________________
7. In what country do you live? _______________________________________
8. In which state or territory is your primary residence? _________________
9. What is your highest level of completed formal education? (circle one)

    - High School or equivalent
    - Associates degree (2-year) or equivalent
    - Bachelors degree (4-year) or equivalent
    - Masters degree or equivalent
    - M.D./J.D/Ph.D or equivalent

10. What is your primary field of profession? (circle one)
    (the following list is from the U.S. Department of Labor)

    - Arts, Design, Entertainment, Sports and Media
    - Building and Grounds Cleaning and Maintenance
    - Community and Social Services
    - Construction and Extraction
    - Education, Training and Library
    - Engineering, Architecture and Surveyors
    - Farming, Fishing and Forestry
    - Financial
    - Food Preparation and Serving Related
    - Health Care Practitioners and Technical
    - Installation, Maintenance and Repair
    - Legal
    - Life, Physical and Social Science
    - Management
    - Mathematical and COmputer Scientists
    - Office and Administrative Support
    - Personal Care and Service
    - Production
    - Protective Service
    - Sales and Related
    - Transportation and Mateiral-Moving
    - Other (please specify) ________________

11. What is your level of astronomy experience? (circle one)
    - Novice with very basic astronomy experience
    - Intermediate level
    - Advanced level but not in a professional capacity
    - Professional astronomer, astrophysicist, etc.

12. What do you consider was the *first* year you were active in the AAVSO? (ex: 1975, 1990, etc.)
_____________________________________________________________________
13. What was the main reason you became active in the AAVSO at that time?
_____________________________________________________________________
14. How did you hear/learn about the AAVSO? ___________________________
15. If you are active in the AAVSO right now, what is your current main interest? _________
16. If you are not currently active in the AAVSO, what is the main reason? _______________
17. Are you an active variable star observer? (circle one)   Yes   No
18. Of all the observing techniques you may use, what is the magnitude of the faintest positive variable star observation that you routinely make (do not count fainterthans)? ___________
19. How do you make those faint observations? (check one)
    ______ Visual          ______ CCD/DSLR/PEP/Other Digital system

20. If you are a CCD/DSLR/PEP observer, did you begin as a visual observer who migrated to CCD/DSLR/PEP or did you begin as a CCD/DSLR/PEP observer? (check one)
__I began making variable star observations as a visual observer who migrated to CCD/DSLR/PEP/etc.
__I began making variable star observations as a CCD/DSLR/PEP/etc. observer.

21. Select the following observing techniques that you actively use (circle as many that apply):
   - Naked eye
   - Binoculars
   - DSLR
   - Telescope visual
   - Telescope CCD
   - Telescope PEP
   - Spectroscopy
   - Sunspot counting
   - SID monitoring
   - AAVSONet or other robotic systems
   - Other (please specify):

22. Select the following nonobserving activities in which you are currently active (circle as many that apply):
   - AAVSOnet hardware or software support
   - Actively contributing to a committee, division or section
   - Chart/Sequence Development
   - Data archiving
   - Data mining
   - Financial support (beyond membership fees)
   - Mentoring
   - Programming
   - Public Speaking and Teaching
   - Have served on Council at any point in time
   - Volunteering at HQ
   - Writing
   - Other (please specify):

23. Please rank the types of objects you are most interested in (fill in the circle).
Please choose only one item for each rank. For example, only one object should be assigned a "1", only one object should be assigned a "2", etc.

1 = Most Interested     5 = Neutral     9 = Least Interested

| | 1 | 2 | 3 | 4 | 5 | 6 | 7 | 8 | 9 |
|---|---|---|---|---|---|---|---|---|---|
| Rotational Variables | ○ | ○ | ○ | ○ | ○ | ○ | ○ | ○ | ○ |
| Cataclysmic Variables (Dwarf Novae, Novalike, etc.) | ○ | ○ | ○ | ○ | ○ | ○ | ○ | ○ | ○ |
| Non-stellar Objects | ○ | ○ | ○ | ○ | ○ | ○ | ○ | ○ | ○ |
| Young Stellar Objects | ○ | ○ | ○ | ○ | ○ | ○ | ○ | ○ | ○ |
| Eclipsing Binaries | ○ | ○ | ○ | ○ | ○ | ○ | ○ | ○ | ○ |
| Novae | ○ | ○ | ○ | ○ | ○ | ○ | ○ | ○ | ○ |
| Pulsating Variables (Cepheids, Miras, RR Lyr, etc.) | ○ | ○ | ○ | ○ | ○ | ○ | ○ | ○ | ○ |
| Extragalactic Objects (supernovae, blazars, etc.) | ○ | ○ | ○ | ○ | ○ | ○ | ○ | ○ | ○ |
| The Sun | ○ | ○ | ○ | ○ | ○ | ○ | ○ | ○ | ○ |
| Other (please specify) | | | | | | | | | |

24. How many AAVSO meetings have you attended? ______________________________
25. If 1 or more, in what year did you last attend an AAVSO meeting? _______________
26. Have you been an author or coauthor of a paper submitted to an astronomical journal (JAAVSO included—check one)? _____ Yes _____ No
27. Do you have any general comments you'd like to share?